\documentclass[%
 aapm,
 mph,%
 amsmath,amssymb,
 reprint,%
]{revtex4-2}

\usepackage{graphicx}
\usepackage{dcolumn}
\usepackage{bm}
\usepackage{upgreek}
\usepackage{xcolor}
\usepackage{color,soul}

\usepackage[mathlines]{lineno}
\modulolinenumbers[5]

\begin{document}

\preprint{AAPM/123-QED}

\title{Grain Boundary Effects in Dealloying Metals: A Multi-Phase Field Study}

\author{Nathan Bieberdorf}
\author{Mark Asta}
\affiliation{
 Department of Materials Science and Engineering, University of California, Berkeley, CA 94720}

\author{Laurent Capolungo}
\affiliation{Materials Science and Technology Division, Los Alamos National Laboratory, Los Alamos, NM 87545, USA}%

\begin{abstract}
A multi-phase field model is employed to study the microstructural evolution of an alloy undergoing liquid dealloying. The model proposed extends upon the original approach of Geslin et al. \citep{Geslin2015LMD} to consider dealloying in the presence of grain boundaries. The model is implemented using a semi-implicit time stepping algorithm using spectral methods, which enables simulating large 2D and 3D domains over long time-scales while still maintaining a realistic interfacial thickness. The model is exercised to demonstrate a mechanism of coupled grain-boundary migration to maintain equilibrium contact angles with this topologically-complex solid-liquid interface during dealloying. This mechanism locally accelerates dealloying by dissolving the less noble alloy metal from (and rejecting the more noble metal into) the migrating grain boundary, thereby enhancing the diffusion-coupled-growth of the liquid channel into the precursor. The deeper corrosion channel at the migrating grain boundary asymmetrically disrupts the ligament connectivity of the final dealloyed structure, in qualitative agreement with published experimental observations. It is shown that these grain boundary migration-assisted corrosion channels form even for precursors with small amounts of the dissolving alloy species, below the so-called \textit{parting limit}.
\end{abstract}

\maketitle

\section*{\label{sec:introduction}Introduction}
Dealloying of multicomponent metals is known to generate highly topologically complex structures with micron- or nano-scale ligaments and voids \citep{Koger1972, Zheng2018, Vukmirovic2002}. Interest in this process stems from its use as a method to produce nanoporous metals at large scales. Dealloyed nanoporous metals have been shown to possess remarkable catalytic properties, large capacitance, and high mechanical strength when compared to their bulk counterparts \citep{Erlebacher2009, Fujita2012, McCue2016bLMD, Kim2015}. However, these desirable properties can be deleteriously affected by the presence of grain boundaries in the metal, which are known to affect the ligament size and connectivity of these dealloyed structures \citep{Harrison1959,McCue2016bLMD,McCue2016LMD,Joo2022,Song2020LMD,Gaskey2019LMD,Yang2022}.

In general, systems relevant to dealloying processes can be described as a ternary, wherein a metal alloy precursor made up of species ‘A’ and ‘B’ is subjected to corrosion by a fluid dealloying agent ‘C.’ The thermodynamics of these species' interactions is such that they are all relatively soluble in one another, except species A and C which are relatively insoluble. This results in the selective dissolution of B into C, while A organizes into a topologically complex solid structure. Some examples of porous metals synthesis via dealloying include aqueous solution dealloying (e.g. Ag dealloying from AuAg in acid \citep{Erlebacher2001ASD,Lu2007ASD}), liquid metal dealloying (e.g. Ti dealloying from TaTi in liquid Cu \citep{McCue2016LMD}), and, more recently, molten salt dealloying (e.g. Cr dealloying from NiCr in molten salt \citep{Liu2021MSD}).

The key mechanisms by which nanoporous microstructure formation proceeds have been identified by means of numerical and experimental investigations. Early work by Erlebacher et al. \citep{Erlebacher2001ASD} illustrated the important role of spinodal driving forces and diffusion processes within the solid-liquid interface in reorganizing the non-dissolving A-species into ligaments. This promotes dissolution of the AB precursor into a \textit{bicontinuous} structure, made up of two continuous, interpenetrating phases of A-rich solid and C-rich liquid. A phase field model for liquid metal dealloying (LMD) of TaTi precursors in liquid Cu was later developed by Geslin et al. \citep{Geslin2015LMD} based on a parameterization informed by experimental observations. Their model successfully predicted the early stages of dealloying and A-ligament formation, and elaborated on this mechanism of interface-diffusion-coupled-growth of the A (Ta) rich solid and C (Cu) rich liquid into the AB (TaTi) precursor.

Although the microstructure of the solid alloy precursor is expected to play an important role in this dealloying, the detailed mechanisms underlying how features such as grain boundaries may alter this process are less understood. Experimental studies have shown that grain boundaries locally alter dealloying, resulting in a corroded microstructure that is topologically- and chemically-distinct from the bulk lattice \citep{Harrison1959,McCue2016bLMD,Song2020LMD,Gaskey2019LMD,Yang2022}. In one example, liquid metal dealloying (LMD) experiments by McCue et al. \citep{McCue2016LMD} on TaTi precursors showed that corrosion at grain boundaries is significantly deeper and generates thick Ta-blocks that asymmetrically disrupt the ligament connectivity of the final porous structure. More recently, experiments by Joo et al. \citep{Joo2022} investigated the early stages of this phenomenon in FeNi precursors dealloyed by liquid Mg. They also showed that infiltration by the dealloying agent at the grain boundary generates thick dealloyed plates that are asymmetrically attached to one of the parent grains while being separated from the other grain by the liquid melt. Interestingly, at different locations along a corroded grain boundary, the connectivity of these dealloyed plates was seen to switch back and forth between parent grains, resulting in a ``wavy" corrosion channel \citep{Joo2022}.

Mechanistic understanding of these experimental observations \citep{McCue2016LMD, Joo2022} remains incomplete. It has been proposed that the excess energy of grain boundaries (GBs) accelerates dealloying through a mechanism of GB-wetting by the liquid phase \citep{Joo2022}. However, the enhanced kinetics at the GB (both interface mobility and solute diffusivity), should also influence this process by locally altering the diffusion-coupled-growth of the interpenetrating solid and liquid phases. It is critical to understand these mechanisms of grain boundary dealloying corrosion to optimize nanoporous materials synthesis.

The current work aims to understand how the presence of grain boundaries affects local dealloying in a model LMD system. The phase field model by Geslin et al. \citep{Geslin2015LMD} is extended to model an AB bicrystal precursor corroding in liquid C, using the multi-phase field formalism of Steinbach and Pezzolla \citep{Steinbach1999}. The multi-phase field model uses an advanced semi-implicit time integration procedure by Badalassi et al. \citep{Badalassi2003}, which allows for accessing large length- and time-scales while still maintaining realistic interfacial thicknesses. 2D and 3D phase field simulations are used to investigate grain boundary effects on dealloying morphology, and numerical predictions are discussed in the context of experimental findings by McCue et al. \citep{McCue2016LMD} and Joo et al. \citep{Joo2022}. In addition, we discuss how these findings may relate to grain boundary accelerated dealloying seen in metals in molten salt environments \citep{Zhou2020, Yang2022}.

\section*{\label{sec:results}Results}

\subsection*{Dealloyed Microstructure at Grain Boundaries}

\begin{figure*}
  \centering
    \includegraphics[width=1\linewidth]{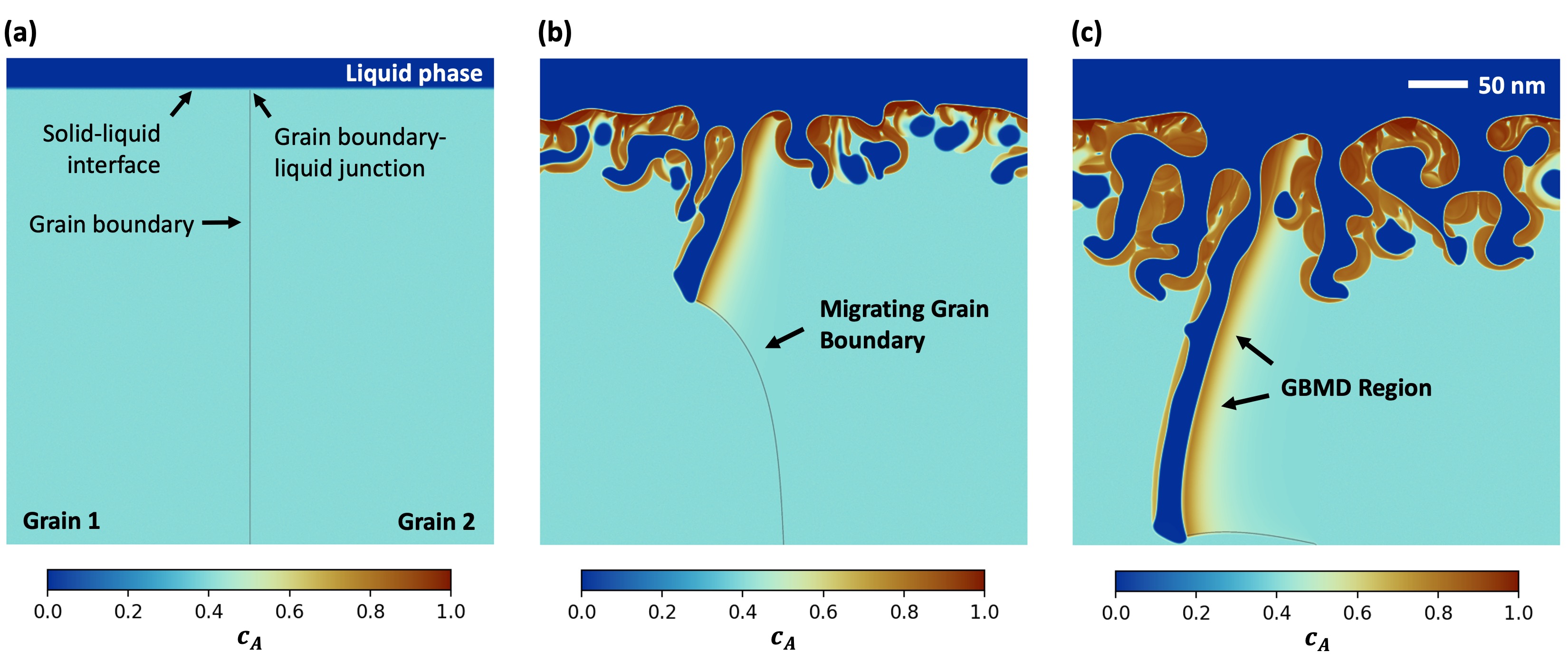}
\caption{Multi-phase field simulations of liquid-metal-dealloyed bicrystal precursor with initial composition $c_{A,0}=40\%$ (a) at the initial condition, (b) after 23 $\upmu$s, and (c) after 72 $\upmu$s. The simulations are colored by the local concentration of the non-dissolving species, A. The migrating grain boundary is marked with a gray band, the solid liquid interfaces for each grain is the surface where $c_A$ abruptly goes to zero. (See Figure \ref{fig:initialcondition} and the Methods section for more details on the initial conditions for the phase field model.) A video of this simulation is provided in Supplemental Materials.}
  \label{fig:bigBicrystal}
\end{figure*}

Figure \ref{fig:bigBicrystal} presents a 2D multi-phase field simulation of a liquid metal dealloyed bicrystal precursor at various time steps. The simulation uses the thermodynamic and kinetic parameterization of TaTi (hereafter AB) exposed to liquid Cu (hereafter C), and grain boundary energies, mobilities and diffusivities set to realistic values as described in the Methods section.  The initial simulation condition and snapshots at 23 and 72 $\upmu$s are shown in Figs. \ref{fig:bigBicrystal} (a)-(c), respectively. These simulations employ a 2D grid of 2048 $\times$ 2048 pixels (409.6 $\times$ 409.6 nm$^2$), initially containing a bicrystal precursor (made up of initial A-concentration $c_{A,0}=40\%$ and B-concentration $c_{B,0}=1-c_{A,0}$) in contact with a liquid phase (made up of pure C). Each image is colored by the local concentration $c_A$, and the gray band tracks the grain boundary. For more details on the simulation cell setup, see the Methods section. Videos of all simulations are provided in the Supplementary Materials.

As shown in Figure \ref{fig:bigBicrystal}(b) and \ref{fig:bigBicrystal}(c), the liquid dealloying agent penetrates the precursor in a bicontinuous manner leading to the formation of an A-rich topologically-complex solid structure. The A-rich solid forms a highly connected ligament network, except at the grain boundary, where a significantly deeper corrosion channel forms and separates the parent grains. Interestingly, the grain boundary appears to couple with the dealloying process by migrating from its original position and imparting a locally-distinct microstructure. Specifically, the grain boundary begins to migrate towards the left and accelerate local corrosion by dealloying B from deep inside (and rejecting A deep into) the bulk lattice through which it sweeps. An A-rich wake forms from the migrating grain boundary, which locally passivates the growing (right) grain at its solid-liquid interface. We label this flat, A-rich structure as the grain boundary migration-assisted dealloying (GBMD) region in Figure \ref{fig:bigBicrystal}(c). The resulting morphology is a bicontinuous network of A-rich ligaments that are disrupted by the larger A-rich block formed by the migrating grain boundary.

\begin{figure*}
  \centering
    \includegraphics[width=1\linewidth]{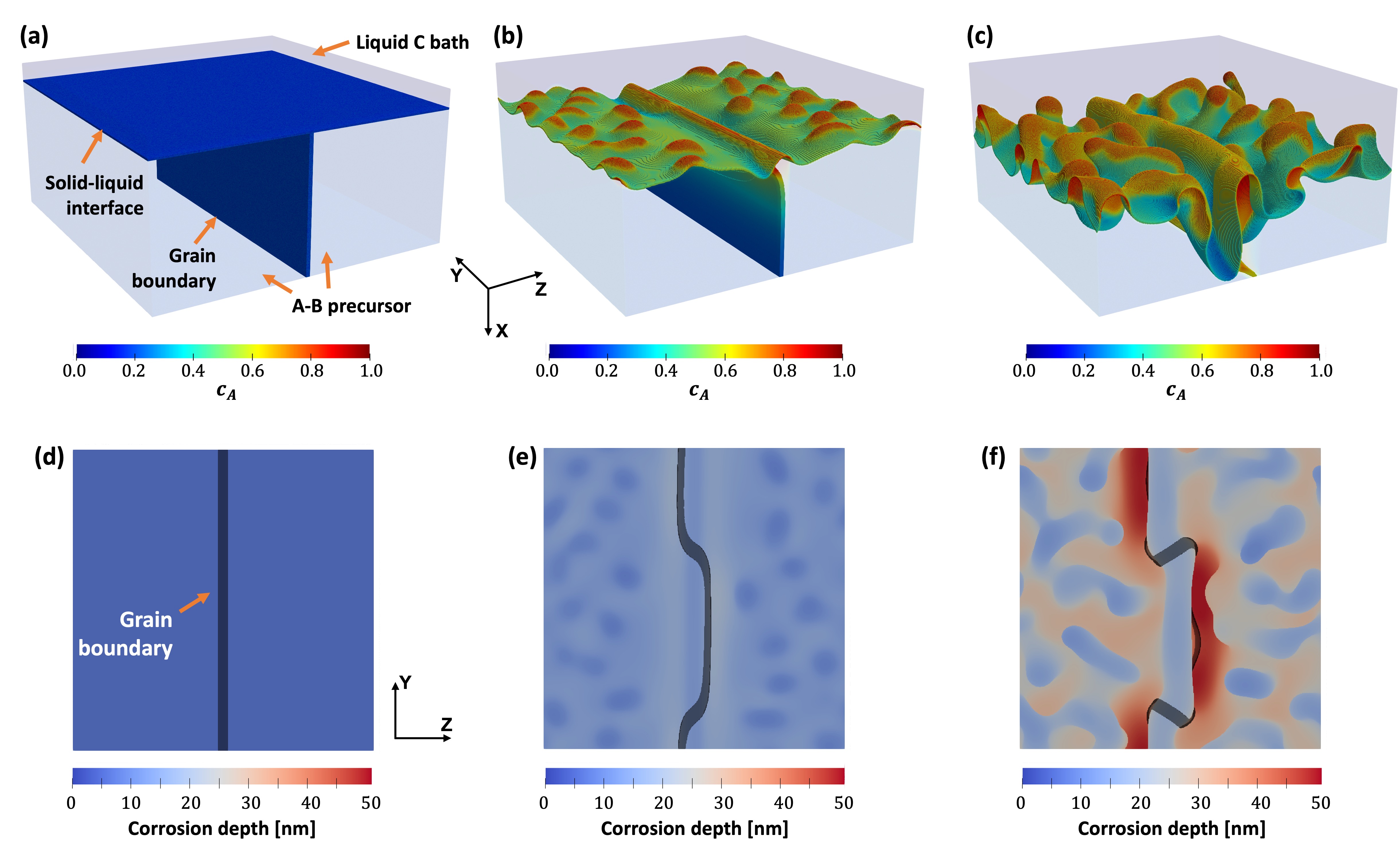}
  \caption{3D multi-phase field simulation of a liquid-metal-dealloyed bicrystal precursor with initial A-composition $c_{A,0}=20\%$ at the initial condition in (a), after 400 ns in (b), and after 950 ns in (c). In these 3D views, only the solid-liquid interface and grain boundary are filled in and colored by the concentration of species A. Topological views looking down (in the positive X-direction) on the solid-liquid interface are shown in the bottom row (d), (e), and (f), with each topological view corresponding to the same time step as the 3D view just above it. The topological maps are colored by the corrosion depth in the positive X-direction. The 3D- and topological-view videos for this simulation are provided in Supplemental Materials.}
  \label{fig:3D}
\end{figure*}

3D simulations are necessary to explore the dealloying corrosion morphology across the solid-liquid interfacial plane, both near and away from the grain boundary-liquid line. Figure \ref{fig:3D} shows successive snapshots from a 3D multi-phase field simulation of a bicrystal precursor with initial A-concentration $c_{A,0}=20\%$ undergoing liquid metal dealloying. In the top row of Figure \ref{fig:3D}, 3D views are shown, where only the interfaces (solid-liquid and grain boundary) are filled in and colored by the concentration of species A. In the bottom row of Figure \ref{fig:3D}, the topological views looking down at the solid-liquid interface for each snapshot are shown, colored by the corrosion depth. In the topological views, the location of the diffuse grain boundary-liquid line is shown by the dark grey band. These simulations employ 256 $\times$ 512 $\times$ 512 voxels (51.2 $\times$ 102.4 $\times$ 102.4 nm$^3$) in the X-, Y-, and Z-directions respectively. 

In the very early stages of the dealloying in Figure \ref{fig:3D}(b) and \ref{fig:3D}(e), A-rich solid nodes begin to set up and the grain boundary begins to migrate, as was seen in early stages of 2D simulations. These 3D simulation snapshots provide additional insight, showing that, away from the grain boundary, these A-rich nodes form as three-dimensional mounds. Meanwhile, along the migrating grain boundary a single two-dimensional ridge forms. Examination of Figure \ref{fig:3D}(e) reveals that the grain boundary is actually migrating in two opposite directions simultaneously – in some locations it has migrated in the positive Z-direction, while in other locations it has migrated in the negative Z-direction. 

As the corrosion continues, the A-rich mounds away from the grain boundary grow and form three-dimensional A-rich ligaments. At the grain boundary, the two-dimensional A-rich ridge persists while corrosion channels rapidly penetrate the precursor, separating each grain’s ligament network from one another. As was the case in 2D simulations, these GBMD corrosion channels form in the direction the grain boundary is locally migrating.

\subsection*{Grain Boundary Migration-Assisted Dealloying Mechanism}

\begin{figure*}
  \centering
    \includegraphics[width=1\linewidth]{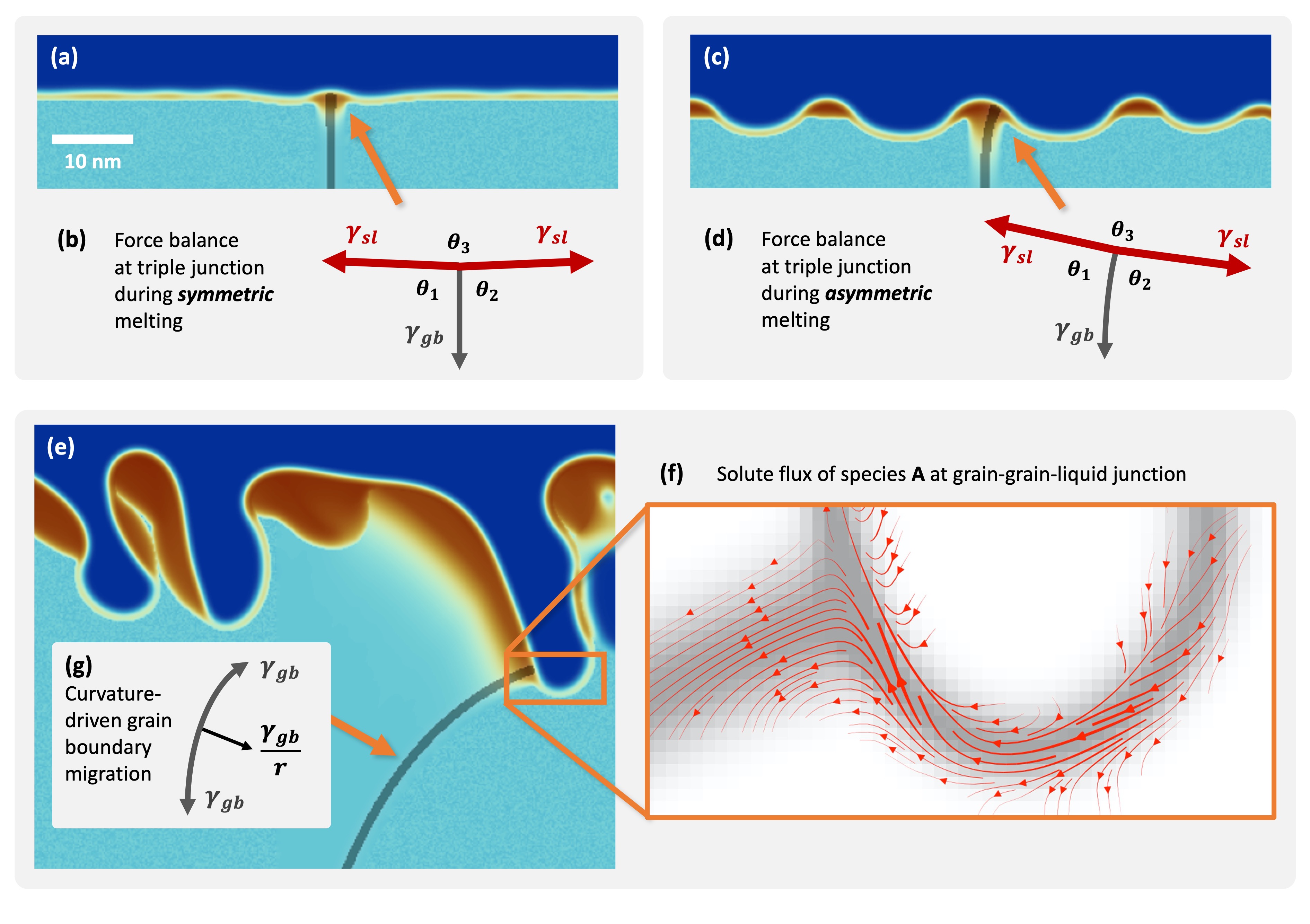}
  \caption{Mechanism of grain boundary migration assisted dealloying, (a) initial force balance of interfaces at the grain boundary-liquid junction when the corrosion front is relatively symmetric about the grain boundary axis shown in the phase field simulation, and (b) shown schematically, (c) grain boundary-liquid junction migration to achieve force balance during in the presence of an asymmetric corrosion front seen in the phase field simulation, and (d) illustrated schematically, (e) later stages of grain boundary migration driven by (f) the flux of species A out of the solid liquid interface down into the grain boundary, and (g) due to curvature forces on the grain boundary.}
  \label{fig:mechanism}
\end{figure*}

The multi-phase field simulation results herein are used to understand in detail the mechanism underlying the coupling between dealloying and grain boundary migration. This mechanism of grain boundary migration-assisted dealloying, or ``GBMD," is driven by the relative energetics of the grain boundary and solid-liquid interface and is reinforced by the solute flux along these interfaces at their junction. Figure \ref{fig:mechanism} illustrates the GBMD mechanism for a bicrystal with $c_{A,0}=35\%$. (For a full description of this dealloying process in the absence of grain boundary effects, the reader is referred to Geslin et al. \citep{Geslin2015LMD}, and only the salient features and grain boundary effects are described herein). Generally throughout the dealloying process, the solid-liquid interface becomes A-rich as B is selectively dissolved from the precursor. Since diffusive transport within the bulk lattice is negligibly small, dealloying is confined to within the solid-liquid interface, except at the grain boundary-liquid junction where diffusion along the grain boundary promotes B-dissolution / A-rejection from / to deeper within the precursor solid alloy. This is shown in Figure \ref{fig:mechanism}(a) at an early stage of the dealloying process, where it can be seen that the dissolution is relatively uniform across the solid liquid-interface except at the grain boundary-liquid junction where it is slightly faster. We note here that, in the present simulations, the grain boundary does not simply 
melt faster than the bulk solid, as one might assume. Rather, since the grain boundary dealloys faster, it builds up in species A, which actually raises the local melting temperature, as it is higher for species A than B. 

At this early stage, the dealloying remains symmetric - from left to right - about the vertical grain boundary. Therefore, the dihedral angles in each grain also remain symmetric ($\theta_1 \approx \theta_2$), with their magnitude determined by the excess interfacial grand potential of the grain boundary, $\gamma_{gb}$, and of the solid-liquid interface, $\gamma_{sl}$. This is schematically shown in Figure \ref{fig:mechanism}(b). The relative energetics of each interface depend on the prescribed values of the grain-boundary and solid-liquid interfaces for the pure elements ($\sigma_{gb}$ and $\sigma_{sl}$ in Eq. \ref{eq:phaseEvolution}) used to parameterize the multi-phase field model, as well as the local chemistry that modifies these interfacial free energies relative to their values for the pure elements. With regards to chemistry, the high concentration of solid-A mixing with liquid-C is seen to significantly increase the solid-liquid interfacial energy in the model, due to their large miscibility gap. This leads to a large value (near $\pi$) for the liquid dihedral angle, $\theta_3$, as seen in Figure \ref{fig:mechanism}(a) and \ref{fig:mechanism}(b).

The dealloying front quickly becomes B-depleted / A-rich to the point that the solid-liquid interface develops a composition that is within the unstable region of the chemical free energy in Eq. \ref{eq:freeEnergyChemical}. This leads to spinodal instabilities for unmixing within the interface, causing segregation of A-colonies from the penetrating C. Importantly, interfacial solute diffusion processes are active enough to enable this A-C segregation. As these A-rich structures continuously set-up and coarsen laterally along the solid-liquid interface, the symmetry about the grain boundary eventually breaks. One example of this is shown in Figure \ref{fig:mechanism}(c), where the topology of the A-rich solid structure has ``tilted" slightly so that it slopes down-right at the grain boundary-liquid junction. To reestablish force balance, the grain boundary-liquid junction and grain boundary migrate (towards the right) until $\theta_1 \approx \theta_2$, as illustrated in Figure \ref{fig:mechanism}(d). We note here that the direction of this asymmetry is random, and grain boundary migration is equally likely to initiate towards the right or the left.

Once this migration is initiated, it is generally reenforced through two coupled mechanisms. First, as the grain boundary moves with the triple junction, it migrates through the bulk lattice, continuously providing new, non-dealloyed regions along which to reject A and dealloy B. This results in the A-rich wake following the grain boundary migration path, seen again here in Figure \ref{fig:mechanism}(e). A snapshot of this A-rejection into the grain boundary is shown in Figure \ref{fig:mechanism}(f), where the local flux of species A is represented by the red streamlines and the gray regions show the interfaces. It can be seen that A - being immiscible with the liquid C and immobile in the bulk lattice - is confined to diffuse along the solid-liquid and grain boundary interfaces. Along the solid-liquid interface, spinodal driving forces drive the ``uphill" diffusion of A towards adjacent A-rich structures. As discussed in Geslin et al. \citep{Geslin2015LMD}, this is essential to the diffusion-coupled-growth of interpenetrating A-rich solid and C-rich liquid phases into the AB-solid precursor - where the rate of diffusion-coupled-growth is controlled by the rate at which species A can be ``evacuated" from the infiltrating C-liquid channel. An essential feature of GBMD is that the grain boundary provides an additional pathway along which to reject species A, thereby locally enhancing this diffusion-coupled-growth.

The second reenforcing mechanism, grain boundary straightening, occurs simultaneously. As the grain boundary-liquid junction drags the grain boundary, significant curvature is induced at locations deeper inside the solid, as shown in Figure \ref{fig:mechanism}(g). If the grain boundary interface mobility is sufficient, capillary forces will move the grain boundary to reduce its curvature.  In the simulations, the high interfacial energy of the solid-liquid interface that arises when it is enriched by species A enforces the grain boundary-liquid triple point to be constrained to maintain force balance. As a result, the grain boundary can only straighten by migrating into non-dealloyed regions of the metal, allowing it to take on more A from the solid-liquid interface. (It is noted here that the “continuous-gradient” boundary conditions, discussed in Methods, essentially pin the grain boundary to its initial position at the bottom edge, and in a real system the grain boundary would migrate to a greater extent than the simulations show here).

\subsection*{Grain Boundary Dealloying and the Parting Limit}

\begin{figure}
\centering
  \includegraphics[width=0.9\linewidth]{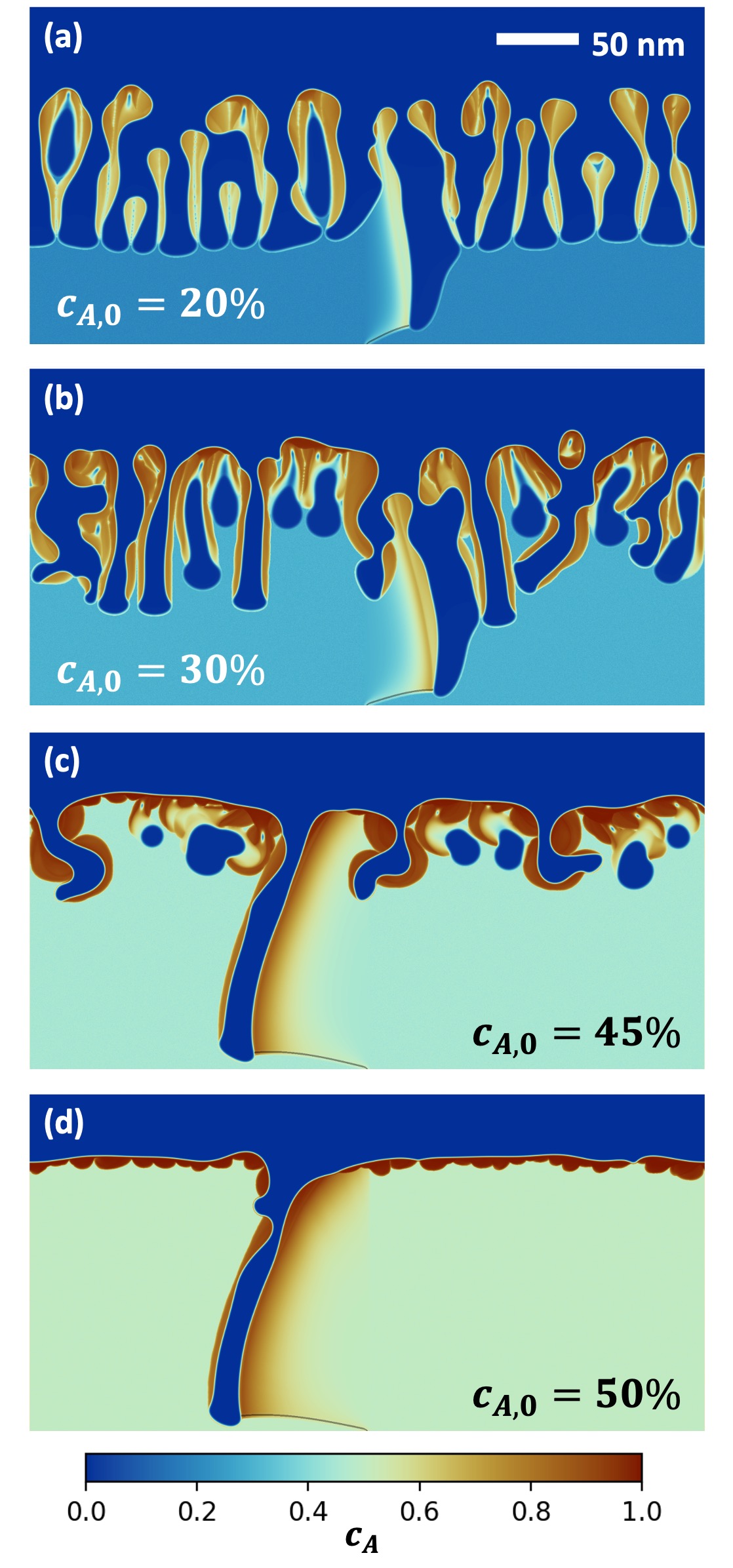}
\caption{Multi-phase field simulations of liquid-metal-dealloyed bicrystal precursors with initial composition (a) $c_{A,0}=20\%$, (b) $c_{A,0}=30\%$, (c) $c_{A,0}=45\%$, (d) $c_{A,0}=50\%$. The simulations are colored by the local concentration of the non-dissolving species, A. The migrating grain boundary is marked with a gray band, the solid liquid interfaces for each grain is the surface where $c_A$ abruptly goes to zero. Videos of these simulations are provided in Supplemental Materials.}
\label{fig:sensitivityComp}
\end{figure}

The degree to which this GBMD mechanism influences the corrosion morphology and depth depends strongly on the initial composition of the precursor. This is illustrated in Figure \ref{fig:sensitivityComp}, which shows dealloyed bicrystal precursors with varying A-content, increasing from $c_{A,0}=20\%$ in Figure \ref{fig:sensitivityComp}(a), $c_{A,0}=30\%$ in Figure \ref{fig:sensitivityComp}(b), $c_{A,0}=45\%$ in Figure \ref{fig:sensitivityComp}(c), up to $c_{A,0}=50\%$ in Figure \ref{fig:sensitivityComp}(d). For comparison, single crystal precursors across this composition range are provided in Supplemental Figure S1.

Precursors with relatively low A-content, $c_{A,0}=20\%$ in Figure \ref{fig:sensitivityComp}(a) and $c_{A,0}=30\%$ in Figure \ref{fig:sensitivityComp}(b), dealloy to form somewhat lamellar structures with alternating A-rich solid and C-rich liquid phases. The diffusion-coupled-growth of each phase is generally sustained since there is a moderately small amount of A to remove from the infiltrating C-channels. The grain boundary accelerates this effect, via the GBMD mechanism illustrated in Figure \ref{fig:mechanism}, leading to a deeper corrosion channel near the migrating grain boundary in both Figure \ref{fig:sensitivityComp}(a) and \ref{fig:sensitivityComp}(b), which separates the ligament networks forming within each grain. 

As the concentration is increased to $c_{A,0}=45\%$ in Figure \ref{fig:sensitivityComp}(c), a more-connected network of thicker A-rich ligaments forms. Since a higher nominal amount of species A must be removed away from the penetrating C-liquid, diffusion-coupled-growth of these channels is significantly slowed, or even broken-down in some cases. And so, away from the grain boundary, interpenetrating A-rich solid and C-rich liquid phases tend to follow more tortuous pathways into the precursor. However, since the fast-diffusing grain boundary locally assists this diffusion-coupled-growth, a singular C-rich channel rapidly penetrates at the migrating grain boundary, forming a linear pathway into the precursor similar to the lamellar channels forming in precursors with lower A-content.

At the highest A-concentration considered, $c_{A,0}=50\%$ in Figure \ref{fig:sensitivityComp}(d), diffusion-coupled-growth of C-channels into the precursor is not possible away from the grain boundary. Indeed, in the single-crystal simulations in Supplemental Figure SM1(d), the phase field model predicts that, in the absence of the grain boundary, the precursor with $c_{A,0}=50\%$ will not undergo bicontinuous dealloying. This high nominal A-content saturates the interface after some initial dealloying, thereby forming a passivating A-rich phase that prevents further dissolution of the precursor. These precursors are below the model's predicted \textit{parting limit}, i.e. the requisite concentration of species B to fully dealloy the precursor. 

However, as shown in Figure \ref{fig:sensitivityComp}(d), a singular dealloying channel can still invade the precursor by the mechanism of GBMD. The result is a deep dealloying channel at the migrating grain boundary but a relatively passivated corrosion front elsewhere. To clarify, the model does not predict that the grain boundary will lower the parting limit, since this singular infiltration by the dealloying agent does not lead to the formation of a fully-dealloyed A-rich structure. Rather, these results show that for a system below the parting limit, the GBMD mechanism can still lead to the fast infiltration at grain boundaries.

\section*{\label{sec:discussion}Discussion}

The predicted morphologies resulting from GBMD can be compared to previous experimental studies of grain boundary effects in precursors undergoing liquid metal dealloying by McCue et al. \citep{McCue2016LMD,McCue2016bLMD} and Joo et al. \citep{Joo2022} (albeit at time- and length-scales much earlier and smaller than the experiments). In these experiments, the dealloying within grains maintains a uniform depth, while at grain boundaries the dealloying is generally much faster/deeper (see Figure 8 of McCue et al. \cite{McCue2016LMD} and Figure 1 of Joo et al. \citep{Joo2022}). The thick, plate-like dealloyed structure that results near the migrating grain boundary in Figure \ref{fig:bigBicrystal}(c) of the present work qualitatively matches the thick Ta-rich blocks shown in Figure 8(b) of McCue et al. \cite{McCue2016LMD} and also the Fe-rich ``abnormal ligaments" illustrated Figure 2(c) and Figure 3(c) of Joo et al. \citep{Joo2022}. As seen in the experiments and here in Figure \ref{fig:bigBicrystal}(c), this A-rich plate only connects to the ligament network within one of the parent grains, and is separated from the other grain by the widening liquid melt - which continues to penetrate deeper into the sample along the migrating grain boundary.

The SEM image provided in Figure 1(c) of McCue et al. \citep{McCue2016bLMD} nicely shows a 3D view of a dealloyed TiTa precursor near a grain boundary, which allows comparison with 3D multi-phase field predictions in Figure \ref{fig:3D} of the present work. In both the experimental and the simulation results, dealloying corrosion forms three-dimensional ligament networks within the grains and two-dimensional ridges along grain boundaries. Further, both experiment and simulation show that the transition from three-dimensional to two-dimensional corrosion morphology is abrupt.

In Joo et al. \citep{Joo2022}, a ``waviness" of several of the dealloyed grain boundaries was illustrated, wherein the connectivity of the flat dealloyed plates periodically switches back and forth between parent grains. We also note this waviness is seen along some of the grain boundaries in Figure 8 of McCue et al. \citep{McCue2016LMD}. In the present work, Figure \ref{fig:3D} shows that the stochastic nature of GBMD can drive grain boundary migration in opposite directions simultaneously. We postulate that this could lead to a wavy dealloyed structure similar to experiments, since in the locations along the Y-axis in Figure \ref{fig:3D} where the grain boundary migrates left (-Z direction), GBMD would impart a flat dealloyed plate along the right (+Z) grain, and vice-versa. We further postulate that a migrating grain boundary could periodically switch migration directions along the X-axis in Figure \ref{fig:3D}, perhaps due to an intermittency in the activation of the GBMD mechanism. Further, potentially larger, simulations would be helpful to explore this phenomenon in more detail.

Although the qualitative agreement between the present simulations and experimental observations described above is noteworthy, detailed quantitative comparisons with experimental results from McCue et al. \citep{McCue2016LMD,McCue2016bLMD} and Joo et al. \citep{Joo2022} would require an understanding of how the numerical predictions herein will evolve of time- and length-scales more consistent with experiments. These dealloyed structures coarsen predominantly via surface diffusion processes \citep{Erlebacher2011,McCue2016LMD,McCue2016bLMD,Geslin2015LMD,Liu2021MSD}, and, away from the grain boundary, ligaments have been shown to maintain a similar morphology over different dealloying durations, with only the characteristic length scale of ligaments increasing over time \citep{McCue2016LMD,QianASD2007,Wada2013}. Regardless of the mechanism, ligaments will coarsen to reduce the overall surface area of the structure via mass transport from surfaces with high curvature to surfaces with lower curvature. The large flat structure formed at the migrating grain boundary should, therefore, grow at the expense of the adjacent highly-curved ligaments. The coarsening rate depends on local curvatures \citep{Conyers1950}, and so coarsening near the grain boundary should be faster than at locations far away from the grain boundary, given that the features within grains are more similar in size and curvature. Therefore, we expect these numerically predicted corrosion morphologies herein to persist, or become even more exaggerated, with coarsening over longer time-scales.

The GBMD mechanism described in this work is distinct from the diffusion-induced-grain-boundary migration (DIGM) mechanism discussed by Balluffi and Cahn \citep{Balluffi1981}. In the present work, an interdiffusivity is used which assumes fluxes among the solutes sum to zero and there is no explicit accounting for vacancy fluxes. Future work could explicitly include a vacancy species, and model point defect enrichment along a grain boundary that could drive migration through a DIGM process of grain boundary dislocation climb. In any case, the current work illustrates a mechanism of grain boundary migration that occurs absent consideration of vacancy flux.

In the present work, the fast penetration alongside the grain boundary by GBMD occurs in the absence of a grain boundary wetting condition. However, a wetting condition has recently been hypothesized to be the key feature of the heterogeneous dealloying seen experimentally at grain boundaries \citep{Joo2022}. Specifically, it has been proposed that the grain boundary is quickly replaced by two solid-liquid interfaces, as would be the case if the appropriate thermodynamic condition is met that the excess energy of the solid-liquid interface is less than half that of the grain boundary. For the present model, the solid-liquid interfacial energy is influenced by the large enthalpy of mixing between species A and C. This mixing enthalpy leads to a spinodal driving force within the interface, which is an essential characteristic of these dealloying systems \citep{Erlebacher2009, Geslin2015LMD}. However, this large mixing enthalpy ($\Omega_{AC}$ in Eq. \ref{eq:phaseEvolution}) leads to a high excess interfacial grand potential, $\gamma_{sl}$, for the dealloyed interface between C-rich liquid and A-rich solid. We note that this excess grand potential (integrated through the solid-liquid interface) is much greater than the solid-liquid interfacial free energy for the pure elements ($\sigma_{sl}$) in Eq. \ref{eq:phaseEvolution}. The grain boundary, by comparison, has a much lower excess interfacial grand potential, since it resides in an AB alloy with ideal mixing thermodynamics such that compositional enrichment does not change its interfacial energy. This is evident in the contact angles formed at the grain boundary-liquid junction, where the liquid-phase dihedral angle approaches $\theta_3=\pi$. 

Quantitative predictions made using the current model stand to be improved by a refinement of thermodynamic and kinetic parameters for the bulk phases and compositions, as well as for interfaces, junctions, and mixtures. For instance, since solute transport along the grain boundary is key to GBMD, predictions by the current work can be extended by atomistically informed models describing how ground boundary diffusion and mobility vary during dealloying. Further, the model highlights the importance of understanding how the mixing enthalpies vary through the interfacial region between the bulk phases, as this would impact the force balance at the triple junction. For example, a higher solubility between species A and C at the solid-liquid interface, which could be captured herein via a ramping-down $\Omega_{AC}$ when $\phi_1\phi_3 > 0$, would lower the predicted excess grand potential of this interface. Similarly, a driving force for solute segregation at the grain boundary would affect the relative interfacial energies and also the migration phenomenon by potentially adding a solute drag effect. The present model could be used in a future study to explore the relative effect of these driving forces through a parameter sensitivity analysis, which could be used to motivate future experiments and lower length scale simulation studies.

We expect that this mechanism of GBMD could also be active in other dealloying systems, e.g. molten salt dealloying (MSD). Indeed, MSD of NiCr precursors has recently been shown to be a viable method of producing three-dimensional bicontinuous metals \citep{Liu2021MSD}. Similar to LMD, in MSD the dealloying of the less noble element (Cr) coupled with interfacial diffusion processes leads to the formation of a porous metal made up of the more noble element (Ni). Further, MSD has been shown to be accelerated at, or in some cases confined to, grain boundaries in the NiCr precursor \citep{Zhou2020, Yang2022}. Since MSD is conducted at high homologous temperatures, with respect to the alloy precursor, we expect that grain boundaries are sufficiently mobile to activate this GBMD mechanism. Future work could explore this mechanism in MSD through the incorporation of electrochemical driving forces into the current model.

\section*{\label{sec:conclusion}Conclusion}
In this work, an advanced multi-phase field model is presented to study metal dealloying in the presence of grain boundaries. This new model expands upon the original work of Geslin et al. \citep{Geslin2015LMD} to consider grain boundaries by incorporating the multi-phase field formulation by Steinbach and Pezzolla \citep{Steinbach1999}. An advanced semi-implicit time integration algorithm by Badalassi et al. \citep{Badalassi2003}, using spectral methods, is employed, which enables simulating the coupled evolution of an arbitrary number of phases, grains, and conserved species, over large length- and time-scales, while still maintaining a realistic width for interfaces – which is crucial for simulating the fine-scale interfacial curvatures seen in dealloying metals.

Using 2D and 3D simulations, the model is exercised in application to a liquid metal dealloying system (TiTa dealloying in liquid Cu), and we predict phenomena that is qualitatively consistent with the grain boundary effects seen in recent dealloying experiments \citep{McCue2016LMD,McCue2016bLMD,Joo2022}. Through a detailed analysis of the simulation results, we identify a mechanism of grain boundary migration-assisted dealloying (GBMD) that controls the morphology of these dealloyed metals. The key feature of this GBMD mechanism is that it assists the diffusion-coupled-growth of the infiltrating liquid dealloying agent, by providing an additional pathway in the metal precursor along which to reject the more noble element (and dealloy the less noble element). For the deliberate synthesis of nanoporous metals by dealloying, this mechanism is predicted to lead to large blocks of the non-dissolving metal species, which asymmetrically disrupt the connectivity of the final bicontinuous structure. For metals designed to withstand dealloying environments, e.g. with high-concentrations of the non-dissolving species, GBMD can still generate a deep dealloying channel that is expected to severely limit the safe operation of these material systems.

\section*{\label{sec:methods}Methods}
\subsection*{Multi-phase field model}
The phase field model for LMD from Ref. \citep{Geslin2015LMD} is extended to model an AB bicrystal precursor corroding in liquid C, using the multi-phase field formalism of Ref. \citep{Steinbach1999}. The bicrystal precursor is made up of two solid phases (one for each grain) with binary composition $c_{A,0}$ and $c_{B,0}=1-c_{A,0}$. The bicrystal is placed in contact with a liquid phase of pure $c_{C,0}=1$. Within one grain, the non-conserved phase field variable $\phi_1$ takes on the value $\phi_1=1$, while in the other grain $\phi_2=1$. In the liquid phase, the non-conserved phase field variable $\phi_3=1-\phi_1-\phi_2=1$. The diffuse interface between grains where $\phi_1\phi_2>0$ represents the grain boundary, and interfaces where $\phi_1\phi_3>0$ or $\phi_2\phi_3>0$ represent the solid-liquid interfaces for each grain. The diffuse region where $\phi_1\phi_2\phi_3>0$ represents the junction among the two grains the liquid phase.

The phase evolution is given by:
\begin{eqnarray}
    \label{eq:phaseEvolution}
    &&\frac{\partial \phi_\alpha}{\partial t} = \nonumber \\
    &&\sum_{\beta}^{\tilde{N}} \frac{\tilde{M}_{\alpha\beta}}{\tilde{N}}\Bigg[\sum_\gamma^{\tilde{N}}\bigg[\big(\sigma_{\beta\gamma}-\sigma_{\alpha\gamma}\big)\bigg(\nabla^2\phi_\gamma+\frac{\pi^2}{\eta^2}\phi_\gamma\bigg) \bigg] \nonumber \\
    && + \frac{\pi^2}{4\eta}h'\Delta g_{\alpha\beta} \Bigg]
\end{eqnarray}
for independent phases $\alpha=1,2$. In the above, $\tilde{M}_{\alpha\beta}$ and $\sigma_{\alpha\beta}$ are the interfacial mobility and excess free energy for the $\alpha$-$\beta$ interface, $h'=\frac{8}{\pi}\sqrt{\phi_\alpha\phi_\beta}$ is the derivative of an interpolation function between phases, $\Delta g_{\alpha\beta}$ is the chemical free energy difference between phases, $\eta$ is the interface thickness, and $\tilde{N}$ is the number of phases locally present. The difference in energy between bulk solid and liquid phases is $\Delta g_{13}=\Delta g_{23}=\Delta g_{\text{sl}}$, and there is no difference in energy between each grain ($\Delta g_{12}=0$). 

In practice, Eq. \ref{eq:phaseEvolution} is only evaluated at $\alpha$-$\beta$ interfaces where $\phi_\alpha \phi_\beta > 0$ ($\tilde{N}=2$), $\alpha$-$\beta$-$\gamma$ junctions where $\phi_\alpha \phi_\beta \phi_\gamma > 0$ ($\tilde{N}=3$), and at the grid-points adjacent to those interfaces and junctions. The exact form of the phase field dynamics equation is used in Eq. \ref{eq:phaseEvolution}, as opposed to the oft-used ``antisymmetric approximation" \citep{Eiken2006}, because of its ability to resolve equilibrium contact angles at interface junctions \citep{Guo2010}.

The concentration of species $i$ evolves by the continuity equation:
\begin{equation}
    \label{eq:compositionEvolution}
    \frac{\partial c_i}{\partial t} = \nabla \cdot M_{ij}(\{\phi_\alpha\})\nabla\mu_j
\end{equation}
with phase-dependent solute mobility matrix $M_{ij}$, and chemical potential $\mu_j$ of species $j$. In Eq. \ref{eq:compositionEvolution}, the $j$-summation is over independent compositions, $c_A$ and $c_B$. The chemical potential is defined by:
\begin{equation}
    \mu_i = \frac{\partial f_\text{chem}}{\partial c_i}-\sum_{j=A,B,C} \kappa \nabla^2 c_j \frac{\partial c_j}{\partial c_i}
\end{equation}
where $f_\text{chem}$ is the bulk chemical free energy, and $\kappa$ is the energy penalty coefficient for composition gradients.

Following \citep{Geslin2015LMD}, the chemical free energy is defined using a regular solution model:
\begin{eqnarray}
    \label{eq:freeEnergyChemical}
    f_\text{chem} = &&\sum_i \bigg[ \frac{k_\text{B}T}{V_a} c_i \log c_i \bigg] \nonumber + \Omega_{AC} c_A c_C \nonumber \\
    && + h(\phi_3)\Delta g_\text{sl}(\{c_i\})
\end{eqnarray}
The first term is the entropy of mixing, with Boltzmann constant $k_\text{B}$, temperature $T$, and atomic volume $V_a$. The second term is the excess enthalpy associated with mixing elements A and C with factor $\Omega_{AC}$. Following \citep{Geslin2015LMD}, $\Omega_{AC} \gg 0$, which leads to a large miscibility gap between A and C, while all other element pairs mix ideally. The third term contains the difference in chemical energy between the liquid and solid phases, $\Delta g_\text{sl}(\{c_i\})$, weighted by a smooth interpolation function $h(\phi_3)$ where $h=1$ in the solid phases ($\phi_1+\phi_2=1$), and $h=0$ in the liquid phase ($\phi_3=1$). The free energy difference between phases is given as:
\begin{equation}
    \Delta g_\text{sl}(\{c_i\}) = \sum_i {c_i L_i \bigg( \frac{T-T_i}{T_i} \bigg)}
\end{equation}
where $T_i$ and $L_i$ are the melting temperature and latent heat of melting, respectively, for species $i$. The interpolation function is: 
\begin{eqnarray}
    \label{eq:interpolationFcn}
    h(\phi_3)=&&\frac{1}{2}+\frac{2}{\pi} \bigg[ (1-2\phi_3) \sqrt{(1-\phi_3)\phi_3} \nonumber\\
    &&+ \frac{1}{2} \arcsin(1-2\phi_3) \bigg]
\end{eqnarray}

The solute mobility is defined as:
\begin{equation}
    M_{ij}(\{\phi_\alpha\}) = \frac{D(\{\phi_\alpha\})V_a}{k_\text{B}T}c_i(\delta_{ij}-c_j)
\end{equation}
with a diffusivity that varies among solid, liquid, and grain boundary regions:
\begin{equation}
    \label{eq:diffusivity}
    D(\{\phi_\alpha\} = D_l\phi_3 + 16D_{gb}\phi_1^2\phi_2^2
\end{equation}
where $D_l$, and $D_{gb}$ are the diffusivities in the liquid phase and grain boundary region, respectively. Following \citep{Geslin2015LMD}, diffusivity in the solid phase is assumed to be negligibly small relative to the liquid phase and is set to zero. The form of the solute mobility matrix ensures that, in Eq. \ref{eq:compositionEvolution}, Fickian diffusion of Ti is recovered in the liquid Cu. 

At the solid-liquid interface for each grain, Eqs. \ref{eq:phaseEvolution} and \ref{eq:compositionEvolution} reduce to similar forms as used in the dual-phase field model by Geslin et al. \citep{Geslin2015LMD}. The only major difference lies in the choice of interpolation function, $h$, between phases. The interpolation function in Eq. \ref{eq:interpolationFcn}, following \citep{Bhogireddy2015}, and the approximation $h'=\frac{8}{\pi}\sqrt{\phi_\alpha\phi_\beta}$ (where $\alpha$=1,2 and $\beta$=3) in Eq. \ref{eq:phaseEvolution} are selected so that chemical driving forces on $\phi_\alpha$ reach zero when $\phi_\alpha=0$. In the present simulations, this is important for preventing the non-physical wetting of one grain along the other grain’s liquid interface. This triple-junction delocalization could have also been addressed by increasing the energy of the triple-junction, as discussed by Nestler et al. \citep{Nestler2005} and H{\"o}tzer et al. \citep{Hotzer2016}. During development of the present model, different thermodynamic descriptions for the grain boundary-liquid junction were explored, including using the ``anti-symmetric" approximation for the multi-phase field equations \citep{Eiken2006} and triple-point energy manipulation. In all cases, the grain boundary migration-assisted dealloying mechanism remained active, suggesting that the choice of triple-junction treatment does not qualitatively affect the results presented herein.

The grain boundary mobility is $\tilde{M}_{12}=1.2$ m/(s$\cdot$GPa), and grain boundary diffusivity is $7\times10^{-10}$ m$^2$/s. The prescribed interface thickness is $\eta=4$ nm. The grain boundary current (i.e. the 1D integral of the diffusivity across the 1-2 interface) is therefore $D_{gb}\cdot 3\eta/8=10^{-18}$ m$^3$/s. These kinetic parameters are within realistic ranges for varying grain boundary types \citep{Molodov1998, Furtkamp1998, Huang1999, Herzig2005}. The composition-gradient energy-penalty is $\kappa=15.0\ \text{eV}/\text{nm}$. All the simulations presented in the main text use a grain boundary energy $\sigma_{12}=0.39$ J/m$^2$, which is chosen to be higher than the solid-liquid interfacial energy ($\sigma_{13}=\sigma_{23}=0.2 $ J/m$^2$) from \citep{Geslin2015LMD}, but not so high as to enforce a grain boundary wetting condition. This grain boundary energy is reasonable for the elevated temperature at which liquid metal dealloying is conducted \citep{Foiles2010,Cheng2018}. All other material properties are taken from directly from Geslin et al. \citep{Geslin2015LMD}. All thermodynamic and kinetic parameters for the multi-phase field model are given in Table \ref{tab:parameters}. For clarity, in Table \ref{tab:parameters} the interfacial parameters for interactions between phases 1 and 2 are denoted ``gb" since they correspond to the grain boundary, and interfacial parameters between phases 1 and 3 and between phases 2 and 3 are denoted ``sl" for the solid-liquid interface.

\begin{table}
    \caption{Thermodynamic and kinetic parameters used in the multi-phase field simulations.}
    \centering
    \begin{tabular}{c|c}
        Parameter & Value \\
        \hline
        $T$ & 1775 K \\
        $\sigma_{gb}$ & 0.39 $\text{J}/\text{m}^2$ \\
        $\sigma_{sl}$ & 0.2 $\text{J}/\text{m}^2$ \\
        $\eta$ & 4 $\times 10^{-9} \text{m}$ \\
        $\kappa$ & 2.4 $\times 10^{-9} \text{J}/\text{m}$ \\
        $L_A$ & 2.82 $\times 10^{9}\text{J}/\text{m}^3$ \\
        $L_B$ & 1.89 $\times 10^{9}\text{J}/\text{m}^3$ \\
        $L_C$ & 1.84 $\times 10^{9}\text{J}/\text{m}^3$ \\
        $T_A$ & 3290 K \\
        $T_B$ & 1941 K \\
        $T_C$ & 1358 K \\
        $V_a$ & 0.01 $\times 10^{-27} \text{m}^{3}$ \\
        $\Omega_{AC}$ & 1.44 $\times 10^{10} \text{J}/\text{m}^{-3}$ \\
        $\tilde{M}_{gb}$ & 1.2 $\text{m}/(\text{s}\cdot \text{GPa})$ \\
        $\tilde{M}_{sl}$ & 12.0 $\text{m}/(\text{s}\cdot \text{GPa})$ \\
        $D_{gb}$ & $7\times 10^{-10} \text{m}^2/\text{s}$ \\
        $D_{l}$ & $7\times 10^{-9} \text{m}^2/\text{s}$ \\
    \end{tabular}
    \label{tab:parameters}
\end{table}

\begin{figure}
\centering
  \includegraphics[width=1\linewidth]{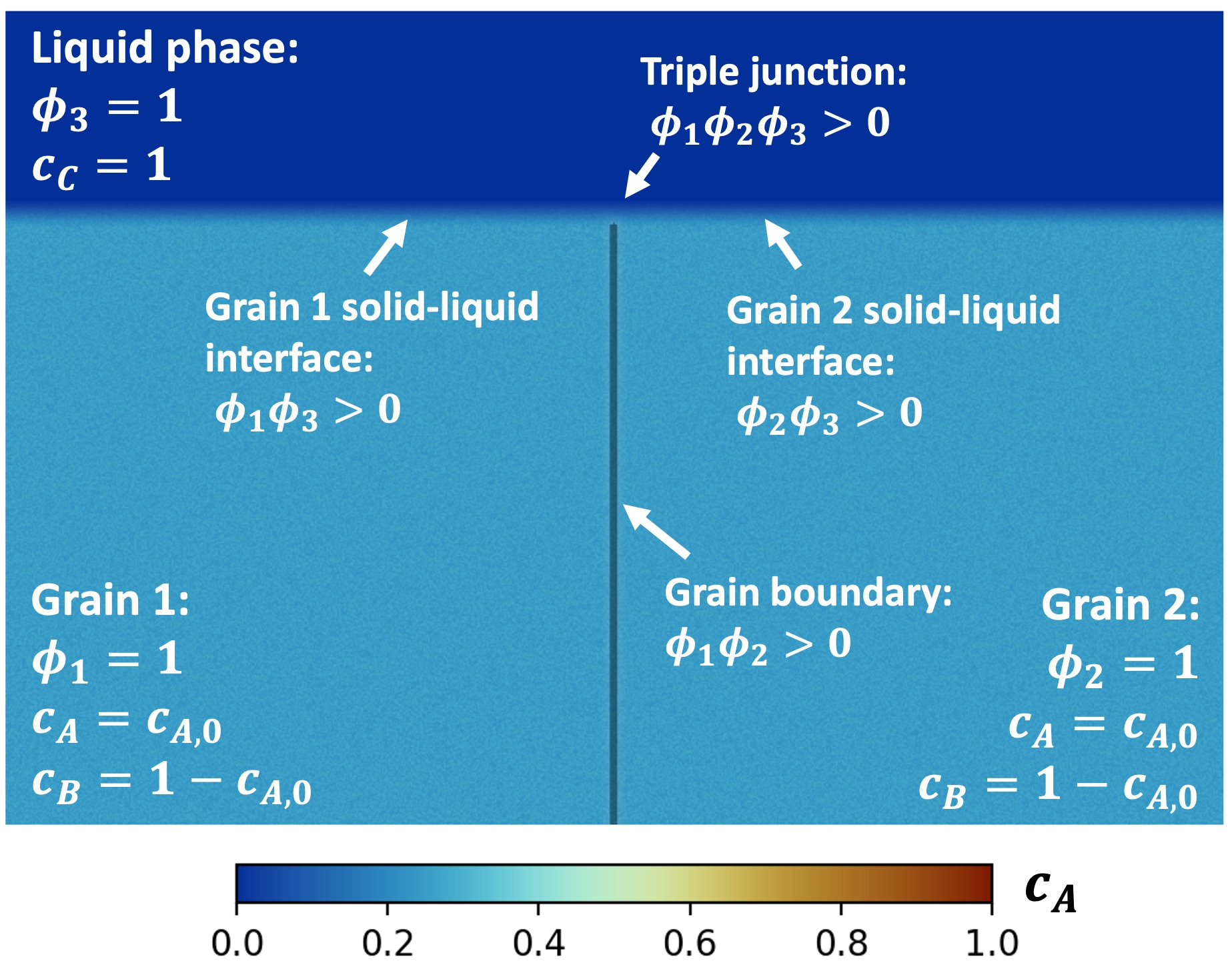}
\caption{Detailed initial condition for multi-phase field simulations of a bicrystal metal precursor with initial AB content $c_{A,0}=0.25,\ c_{B,0}=1-c_{A,0}$, in contact with pure C liquid, zoomed-in at the grain boundary-liquid junction.}
\label{fig:initialcondition}
\end{figure}

\subsection*{Numerical implementation}
A representative simulation cell is shown in Figure \ref{fig:initialcondition}. The simulations are initialized with varying initial alloy compositions $c_{A,0}=0.2$,  $0.3$, $0.4$, $0.45$, and $0.5$ and, following \citep{Geslin2015LMD}, each alloy’s composition is perturbed around its initial value with a uniform white noise of amplitude $\pm$0.025. Dirichlet boundary conditions are enforced on the top of and bottom edges of each simulation. On the left and right edges, periodic boundary conditions are enforced for the conserved species and the non-conserved liquid phase. Periodic boundary conditions on the right and left edges for the non-conserved grains would result a sharp transition between $\phi_1$ and $\phi_2$, which would quickly introduce a second grain boundary. To avoid this, each grain's phase takes on the value of its periodic neighbor at the right and left edges on the simulation cell. That is, $\phi_1$ assumes the value of $\phi_2$, and vice-versa, past the left and right edges when solving Eq. \ref{eq:phaseEvolution}. These ``quasi-periodic" boundary conditions in the horizontal direction are preferred to Neumann (mirror-plane) boundary conditions, so that corrosion channels can freely ``wander" across the left and right edges. (In the single crystal simulations, this not an issue, and regular periodic boundary conditions are enforced for all conserved and non-conserved variables). To avoid boundary effects along the bottom edge, “continuous -gradient” boundary conditions are enforced on phases $\phi_1$, $\phi_2$, and $\phi_3$ following \citep{Guo2010}. This allows the grain boundary to intersect the bottom boundary at non-right angle without penalty.

The phase fields are updated using a forward Euler time integration. However, using forward Euler to update the composition fields would require prohibitively small time-steps to maintain numerical stability. Therefore, the composition fields are updated using the semi-implicit time-stepping procedure following \citep{Badalassi2003}, which uses spectral methods. In the present work, Eq. \ref{eq:compositionEvolution} is computed in real-space, and then $c_i$ and ${\partial c_i}/{\partial t}$ are transformed into frequency-space to semi-implicitly update the composition fields. Discrete sine-transforms are used in the vertical directions to maintain Dirichlet boundary conditions, while discrete real-to-complex transforms are used in the horizontal directions for the periodic boundary conditions. The FFTW library is used to compute transforms \citep{FFTW}, and finite difference operators for derivatives in frequency-space are used to avoid spurious Gibbs oscillations in the composition fields \citep{Berbenni2014}. For more details on the numerical implementation in frequency-space, see Supplementary Materials SM1. In the present work, the semi-implicit time-stepping scheme allows taking time-steps of size $\Delta t = 10^{-3}$ ns, which is approximately 100 times larger when compared to a forward Euler scheme, without loss of stability or accuracy. (Even larger time steps could be used, and still maintain numerical stability, but were found to noticeably change the simulation results). Since numerical integration of Eq. \ref{eq:phaseEvolution} is prone to violating the physical constraint that phases remain on the Gibbs simplex, i.e. $0 \leq \phi_{\alpha} \leq 1$ \citep{Tschukin2017GibbsSimplex}, unphysical phases $\{\phi_\alpha\}$ are projected back onto the Gibbs simplex following \citep{Chen2011GibbsSimplex}. The simulations are discretized with gridpoint spacing $\Delta x$ = 0.2 nm, and Eqs. \ref{eq:phaseEvolution} and  \ref{eq:compositionEvolution} are computed using the isotropic finite difference scheme by Ji et al. \citep{Ji2022}.

Post-processed figures of dealloying precursors were generated using colormaps from Crameri \citep{Crameri2021}.

\section*{Data Availability}
The data produced and analysed in this study are available from the corresponding author on reasonable request.

\section*{Code Availability}
The underlying code developed for this study is not publicly available but may be made available on reasonable request from the corresponding author.

\section*{\label{sec:references}References}
\bibliography{references}

\begin{acknowledgments}
The authors thank Dr. Ian McCue for fruitful discussion of their liquid metal dealloying experiments and results.

This work was supported as part of FUTURE (Fundamental Understanding of Transport Under Reactor Extremes), an Energy Frontier Research Center funded by the U.S. Department of Energy, Office of Science, Basic Energy Sciences. This research used resources of the National Energy Research Scientific Computing Center, a DOE Office of Science User Facility
supported by the Office of Science of the U.S. Department of Energy under Contract No. DE-AC02-05CH11231 using NERSC award BES-ERCAP0020694.
\end{acknowledgments}

\section*{Author Contributions}
N.B., M.A., and L.C. conceived this research. N.B. wrote the simulation code, performed simulations, and analyzed the results with supervision and input from M.A. and L.C. All authors contributed to the writing of the manuscript.

\section*{Competing Interests}
The authors declare no competing interests.

\end{document}


\pagenumbering{arabic} 

\section*{Grain Boundary Effects in Dealloying Metals: A Multi-Phase Field Study}
\section*{Supplementary Materials}

\subsection{Single crystal dealloying simulations}

\renewcommand{\thefigure}{SM1}
\begin{figure}[h]
\centering
\includegraphics[width=1\linewidth]{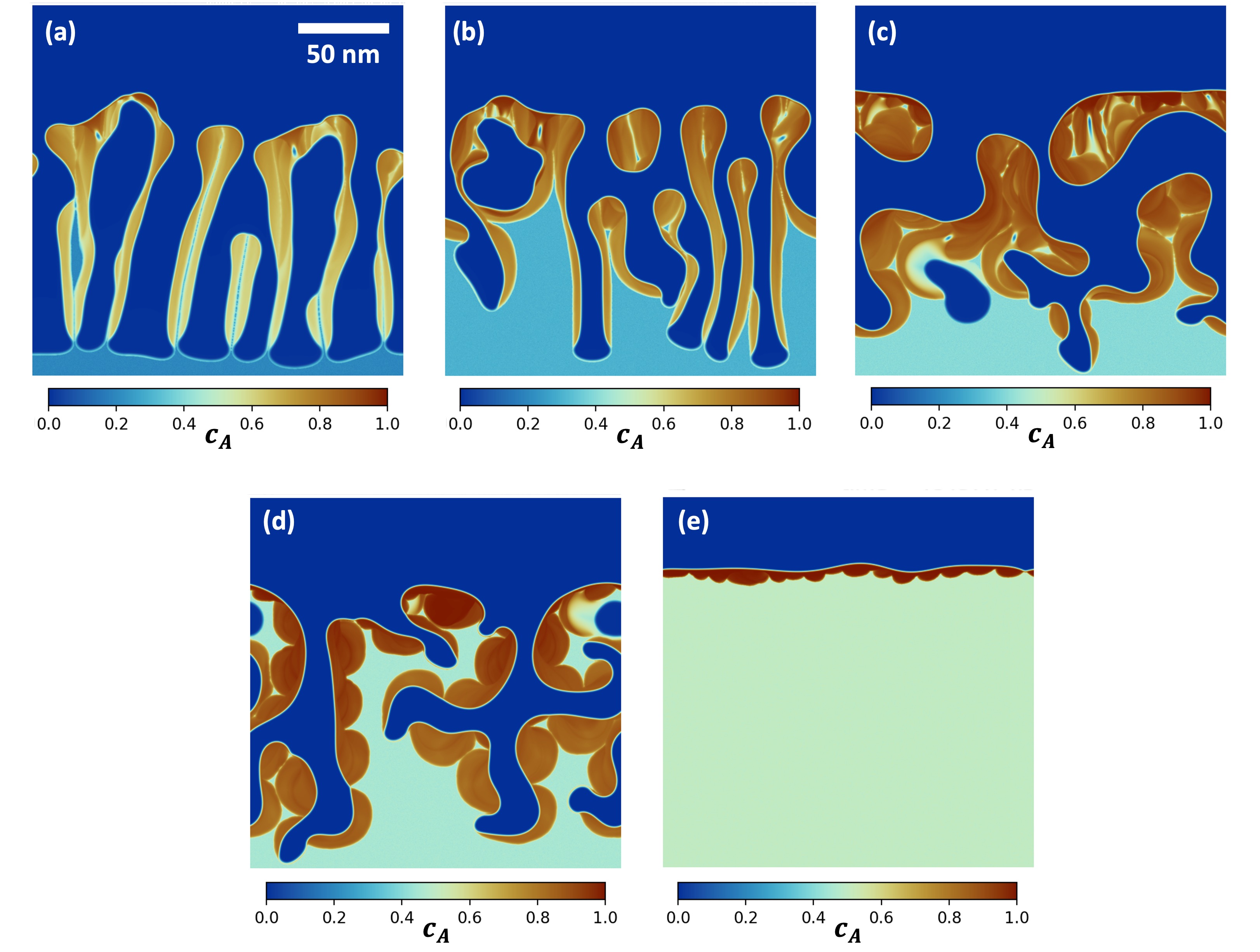}
\caption{Phase field simulations of liquid-metal-dealloyed precursors with initial composition (a) $c_{A,0}=20\%$ after 14 $\upmu$s, (b) $c_{A,0}=30\%$ after 20 $\upmu$s, (c) $c_{A,0}=40\%$ after 74 $\upmu$s, (d), $c_{A,0}=45\%$ after 84 $\upmu$s, (e) $c_{A,0}=50\%$ after 46 $\upmu$s.}
\label{fig:dealoyingVsComposition}
\end{figure}

\clearpage

\subsection{Semi-implicit time integration scheme}

A semi-implicit scheme is used to update the composition in time, based on the multi-step time-evolution equation proposed in Badalassi et al. \cite{Badalassi2003}:
\begin{equation}
\begin{split}
    \frac{ \frac{3}{2}(c_i)^{n+1}_\mathbf{m} - 2(c_i)^{n}_\mathbf{m} + \frac{1}{2}(c_i)^{n-1}_\mathbf{m}}{\Delta t} &= \frac{\lambda}{2}\big[\tau (\nabla^2 c_i)^{n+1}_\mathbf{m} - (\nabla^4 c_i)^{n+1}_\mathbf{m} \big] \\ & + 2\Big((\dot{c}_i)^{n}_\mathbf{m} - \frac{\lambda}{2}\big[\tau (\nabla^2 c_i)^{n}_\mathbf{m} - (\nabla^4 c_i)^{n}_\mathbf{m} \big] \Big) \\ & - \Big((\dot{c}_i)^{n-1}_\mathbf{m} - \frac{\lambda}{2}\big[\tau (\nabla^2 c_i)^{n-1}_\mathbf{m} - (\nabla^4 c_i)^{n-1}_\mathbf{m} \big] \Big)
\end{split}
\label{eq:SM_compEvolution}
\end{equation}
where $(c_i)^{n+1}_\mathbf{m}$ is the composition of species $i$ at gridpoint $\mathbf{m}=\langle m_x,m_y,m_z \rangle$, which is being solved for at next time-step, $n+1$, based on the composition at the current and previous time-steps, respectively $(c_i)^{n}_\mathbf{m}$ and $(c_i)^{n-1}_\mathbf{m}$, and their time-evolution, $(\dot{c}_i)^{n}_\mathbf{m}$ and $(\dot{c}_i)^{n-1}_\mathbf{m}$, which are computed using Eq. 2 of the main text. Along each direction, the gridpoints are indexed $m_j=1,2,...,N_j$, where $N_j$ is the grid-length in the $j$-direction. In Eq. \ref{eq:SM_compEvolution}, $\Delta t$ is the time-step size, and $\lambda$ and $\tau$ are semi-implicit parameters as detailed in \cite{Badalassi2003}. 

Equation \ref{eq:SM_compEvolution} is solved using spectral methods. We define the following discrete Fourier transform:
\begin{equation}
     (\hat{c}_i)^n_\mathbf{k} = \text{DFT}\big[ (c_i)^{n}_\mathbf{m} \big]_\mathbf{k}
\label{eq:ForwardTransform}
\end{equation}
and associated inverse discrete Fourier transform as:
\begin{equation}
     (c_i)^n_\mathbf{m} = \text{DFT}^{-1}\big[ (\hat{c}_i)^{n}_\mathbf{k} \big]_\mathbf{m}
\label{eq:BackwardTransform}
\end{equation}
where $(\hat{c}_i)^n_\mathbf{k}$ is the spectral-space representation of species $i$ at frequency $\mathbf{k}=\langle k_x,k_y,k_z \rangle$, at a given time-step $n$. Along each direction, the discrete frequencies are indexed $k_j=1,2,...,N_j$. The $\text{DFT}$ operator in Eq. \ref{eq:ForwardTransform} is a multi-dimensional discrete Fourier transform, which is chosen based on the boundary conditions in each direction. In the current work, Dirichlet boundary conditions are enforced along the X-direction, and periodic boundary conditions are enforced for the composition evolution along the Y- and Z-directions. This requires using the sine transform (DST-II) along the X-axis and using real-to-complex transforms along the Y- and Z-axes, and so Eq. \ref{eq:ForwardTransform} is fully defined as:
\begin{equation}
\begin{split}
    (\hat{c}_i)^n_\mathbf{k} = 2\sum_{m_z=1}^{N_z} \sum_{m_y=1}^{N_y} \sum_{m_x=1}^{N_x}&(c_i)^n_\mathbf{m} \sin \bigg( \frac{\pi (m_x-\frac{1}{2}) (k_x)}{N_x} \bigg) 
    \\ & \times \exp\bigg( \frac{-2 \pi (m_y-1)(k_y-1) \sqrt{-1}}{N_y} \bigg) 
    \\ & \times \exp\bigg( \frac{-2 \pi (m_z-1)(k_z-1) \sqrt{-1}}{N_z} \bigg) 
\end{split}
\end{equation}
The associated inverse transform uses the inverse sine transform (DST-III), and complex-to-real transforms. Eq. \ref{eq:BackwardTransform} is fully defined:
\begin{equation}
\begin{split}
    (c_i)^n_\mathbf{m} =  \frac{1}{2 N_x N_y N_z}  \sum_{k_z=1}^{N_z} \sum_{k_y=1}^{N_y} & \bigg[(-1)^{m_x-1}(\hat{c}_i)^n_{\langle N_x,k_y,k_z \rangle} + 2\sum_{k_x=1}^{N_x-1} (\hat{c}_i)^n_\mathbf{k} \sin \bigg( \frac{\pi (m_x-\frac{1}{2}) (k_x)}{N_x} \bigg)\bigg] 
    \\ & \times \exp\bigg( \frac{2 \pi (m_y-1)(k_y-1) \sqrt{-1}}{N_y} \bigg) 
    \\ & \times \exp\bigg( \frac{2 \pi (m_z-1)(k_z-1) \sqrt{-1}}{N_z} \bigg) 
\end{split}
\end{equation}
The FFTW library\cite{FFTW} is used to compute these discrete transforms. We note that the definition for the discrete sine transforms imposes the Dirichlet boundary conditions at locations one half-grid-spacing outside of the domain, i.e. at $m_x = 1/2$ and $m_x=N_x + 1/2$. 

The utility of using spectral methods to solve Eq. \ref{eq:SM_compEvolution} is that $(c_i)^{n+1}_\mathbf{m}$ can be solved for directly, since spatial derivatives can be expressed in frequency-space as follows:
\begin{equation}
    \text{DFT}\big[ (\nabla^2 c_i)^{n}_\mathbf{m} \big]_\mathbf{k} = P_\mathbf{k} \cdot (\hat{c}_i)^n_\mathbf{k}
\end{equation}
\begin{equation}
    \text{DFT}\big[ (\nabla^4 c_i)^{n}_\mathbf{m} \big]_\mathbf{k}= Q_\mathbf{k} \cdot (\hat{c}_i)^n_\mathbf{k}
\end{equation}
with 2nd- and 4th-order derivative operators $P_\mathbf{k}$ and $Q_\mathbf{k}$, respectively:
\begin{equation}
\begin{split}
    P_\mathbf{k} = \frac{1}{(\Delta x)^2} \big[ &2\cos(\pi k_x / N_x) \\ 
    + &2\cos(2\pi (k_y-1) / N_y) \\
    + &2\cos(2\pi (k_z-1) / N_z)- 6 \big ]
\end{split}
\end{equation}
\begin{equation}
\begin{split}
    Q_\mathbf{k} = \frac{1}{(\Delta x)^4} \big[ & 2 \cos(2 \pi k_x / N_x) - 8 \cos(\pi k_x / N_x) \\
    + & 2 \cos(4 \pi (k_y-1) / N_y) - 8 \cos(2\pi (k_y-1) / N_y) \\
    + & 2 \cos(4 \pi (k_z-1) / N_z) - 8 \cos(2\pi (k_z-1) / N_z) + 18 \big] 
\end{split}
\end{equation}
where $\Delta x$ is the grid-spacing, which is equal in the X-, Y-, and Z-directions. In the above, we have used finite difference operators, rather than continuous frequency-space derivatives. This choice avoids the spurious numerical Gibbs fluctuations that are known to arise when using continuous frequency-space derivatives on field variables with strong spatial gradients \cite{Berbenni2014}. While phase field model variables are generally smooth in nature, in this work the composition fields do exhibit large spatial gradients at the solid-liquid interface.

Using the above expressions allows writing Eq. \ref{eq:SM_compEvolution} in frequency-space to directly solve for $(\hat{c}_i)^{n+1}_\mathbf{k}$:
\begin{equation}
\begin{split}
    &(\hat{c}_i)^{n+1}_\mathbf{k} = \\ 
    &\frac{ \Big(2 - \Delta t \lambda \big[\tau P_\mathbf{k} - Q_\mathbf{k}\big]\Big) (\hat{c}_i)^n_\mathbf{k} - \Big(\frac{1}{2} - \frac{1}{2}\Delta t \lambda \big[\tau P_\mathbf{k} - Q_\mathbf{k}\big]\Big) (\hat{c}_i)^{n-1}_\mathbf{k} + 2 \Delta t \big(\hat{\dot{c}}_i\big)^n_\mathbf{k} - \Delta t \big(\hat{\dot{c}}_i\big)^{n-1}_\mathbf{k}}{\frac{3}{2}-\frac{1}{2} \Delta t \lambda \big[\tau P_\mathbf{k} - Q_\mathbf{k}\big]}
\end{split}
\end{equation}
which is transformed into real space by using Eq. \ref{eq:BackwardTransform}.

\medskip

\bibliographystyle{plain}
\bibliography{references}